\documentclass[useAMS,usenatbib]{mn2e}

\usepackage{graphicx}
\usepackage{subfigure}
\usepackage{array}
\usepackage{times}
\usepackage{color}
\usepackage{dblfloatfix}

\title[Dynamical stability of the Gliese 581 exoplanetary system]
  {Dynamical stability of the Gliese 581 exoplanetary system}
\author[Zs. T\'oth and I. Nagy]
  {Zs.~T\'oth,$^1$\thanks{Email: zsuzsanna.toth@gmail.com (ZsT); i.nagy@astro.elte.hu (IN)}
  I.~Nagy,$^{2, 3}$\\
  $^1$Department of Geosciences, University of Bremen, Klagenfurter Stra{\ss}e, 
	23359 Bremen, Germany\\
  $^2$Department of Natural Science, National University of Public Service, Hung\'aria k\"{o}r\'ut 9-11, 
	1101 Budapest, Hungary\\
	$^3$Department of Astronomy, E\"{o}tv\"{o}s Lor\'and University, P\'azm\'any P\'eter s\'et\'any 1/A,
	1117 Budapest, Hungary}

\date{Accepted xxxx Month xx. Received xxxx Month xx; in original form xxxx Month xx}

\pagerange{\pageref{firstpage}--\pageref{lastpage}} \pubyear{2013}

\def\LaTeX{L\kern-.36em\raise.3ex\hbox{a}\kern-.15em
    T\kern-.1667em\lower.7ex\hbox{E}\kern-.125emX}

\begin{document}

\label{firstpage}

\maketitle

\begin{abstract}
Using numerical methods we investigate the dynamical stability of the Gliese 581 exoplanetary system. The system is known to harbour four planets (b-e). The existence of another planet (g) in the liquid water habitable zone of the star is debated after the latest analyses of the radial velocity (RV) measurements. We integrated the 4 and 5-planet model of Vogt et al. (AN 333, 561-575, 2012) with initial circular orbits. To characterize stability, the maximum eccentricity was used that the planets reached over the time of the integrations and the LCI and RLI to identify chaotic motion. Since circular orbits in the RV fits seem to be a too strong restriction and the true orbits might be elliptic, we investigated the stability of the planets as a function of their eccentricity. The integration of the circular 4-planet model shows that it is stable on a longer timescale for even an inclination $i = 5^{\circ}$, i. e. high planetary masses. A fifth planetary body in the 4-planet model could have a stable orbit between the two super-Earth sized planets c and d, and beyond the orbit of planet d, although another planet would likely only be stable on circular or near-circular orbit in the habitable zone of the star. Gliese 581 g in the 5-planet model would have a dynamically stable orbit, even for a wider range of orbital parameters, but its stability is strongly dependent on the eccentricity of planet d. The low-mass planet e, which quickly became unstable in eccentric models, remains stable in the circular 4-planet model, but the stable region around its initial semi-major axis and eccentricity is rather small. The stability of the inner planets e and c is dependent on the eccentricity of the Neptune-size planet b. The outermost planet d is far away from the adjacent planet c to considerably influence its stability, however, the existence of a planet between the two super-Earth planets c and d constrains its eccentricity.
\end{abstract}


\section{Introduction}

Based on high-precision radial velocity (RV) measurements acquired using the HARPS \citep{B05, U07, M09} and HIRES \citep{V10} spectrographs, an exoplanetary system was discovered around the red dwarf star Gliese 581 recently. 
With new RV measurements over time and different analyses of the individual and combined RV data sets, the number of planets revealed in the planetary system has varied between three and six: Gliese 581 e, b, c are certain, the planetary signal of Gliese 581 d became questionable in light of a recent analysis \citep{B13}, while the existence of f and g is heavily disputed \citep{AE10b, F11, G11, T11, V12, TdS12, M12, T12, H13}. 

The first four planets (b, c, d, and e) were all discovered in the HARPS data sets by \citet{B05, U07, M09}, respectively. Gliese 581 b is a Neptune-mass planet on a 5.36-day orbit, Gliese 581 c and d are two super-Earth sized planets with orbital periods of 12.9 and 67 days, and Gliese 581 e is a low-mass planet closest to the star with an orbital period of 3.15 days. The super-Earth planets in the system are located on the two edges of the liquid water habitable zone, and while the greenhouse effect of the atmosphere would make planet c too warm and therefore unable to host liquid water, high concentrations of carbon dioxide or other greenhouse gases would be sufficient to keep planet d from freezing out at the cold edge of the habitable zone \citep{S07}. 

Dynamical stability and evolution of the 3-planet system (Gliese 581 b, c and d) of \citet{U07} was investigated by \citet{B08} for different inclinations and was found almost always dynamically stable, even in close to pole-on configurations. The semi-major axes remained extremely stable, the eccentricities underwent only small amplitude variations over the $10^8$ year integration time. In this 3-planet model, both planet c and d had slightly elliptic orbits ($e_c=0.16$, $e_d=0.12$), and they showed significant chaotic behaviour based on their Ljapunov exponents. \citet{ZA09} investigated the dynamical possibility of a fourth planet. A $\sim 2.5 M_{\oplus}$ planet with a semi-major axis between 0.11-0.21 AU (between planet c and d) and eccentricity below 0.25 would not have compromised the overall stability of the system, although the fourth planet was later found inside the orbit of planet b at 0.028 AU. The 4-planet model including the then discovered planet e was integrated by \citet{M09} for $10^8$ years, with the eccentricity of planet e and b fixed at 0, of planet c and d greater than 0 ($e_c=0.17$, $e_d=0.39$). This 4-planet solution remained only stable for $i \geq 30^{\circ}$, because planet e escaped the system after a few Myrs for lower inclinations. Allowing planet e's eccentricity to be 0.1 tightened even more the minimum inclination constraint. The addition of a fifth $\sim 1-10 M_{\oplus}$ planet to this 4-planet model within 0.3 AU quickly introduced dynamical instability to the system, mainly as a consequence of the highly eccentric orbits of the two super-Earth planets c and d \citep{toth11}.

Combining the 4.3-year HARPS set with the 11-year set of HIRES RVs, \citet{V10} announced the discovery of planet f and g orbiting the star with periods of 433 and 36.6 days. Both planet f and g were indicated in the combined RV data sets using a Keplerian fit with (forced) circular orbits. The quality of the fit could be improved only when the eccentricities of the 67-day and 36.6-day planets' orbits were allowed to float, with these two planets being in secular resonance. Gliese 581 g with a mass of $\sim3 M_{\oplus}$ would be a rocky planet in the middle of the habitable zone of the system and might be habitable for a wider range of atmospheric conditions.

Bayesian re-analysis of both the combined and individual datasets by \citet{G11} and \citet{T11} confirmed only four clear planetary signals (planet e, b, c and d) and higher probability for the existence of planet f (with an orbital period of $\sim400$ days). Both studies found that the eccentricities for 3 of the 4 orbits are consistent with zero, the orbit of planet d is elliptical ($e_{d}\simeq0.4$), similarly to the earlier 4-planet solution of \citet{M09}. 

\citet{AE10a} noted how solutions fitting RV data sets with a single-planet eccentric orbit can hide two planets in circular resonant orbits. This is because there is a degeneracy between the resonant and eccentric solutions, as their Keplerian motion equations are identical up to the first order in the eccentricity. In a subsequent study, \citet{AE10b} showed that the first eccentricity harmonic of Gliese 581 d ($\sim33.5$ days) coincides with a yearly alias of the newly reported planet g ($\sim33.2$ days), thus the high eccentricity of planet d can partially absorb the signal from planet g. Nevertheless, based on statistical tests they concluded that the presence of planet g is well supported by the available RV data. \citet{TdS12} similarly concluded that the existence of the 36-day planet g depends on the eccentricity of the 67-day planet d and its detection requires the assumption that all planets are on circular orbits. The signal of planet f was found, but only in the threshold of their confidence level with a period of $\sim455$ days.

\citet{F11} released an additional set of HARPS RV measurements and analysed the then total 7-year data set. Their 4-planet Keplerian-fits, with either fixed or freely floating eccentricities, revealed no significant residual signals after identifying four, therefore no support for the two additional planets g and f. The additional measurements revised the mass of planet d down to 6 $M_{\oplus}$, making a rocky composition more likely. \citet{V12} then re-analysed this data set and warned that allowing the eccentricities to float, and in particular the eccentricity of planet e to rise, leads to instability and therefore to highly unphysical Keplerian models. None of the N-body simulations of the eccentric Keplerian fit of \citet{F11} remained dynamically stable on longer time scales due to high eccentricity of planet e. The 4-planet all-circular interacting model of \citet{V12} on the other hand remained dynamically stable for 20 Myrs. Furthermore, it offered confirmative support for a fifth planetary signal near 32-33 days, which could be planet g at its 36-day yearly alias period.

The recent analyses of \citet{B13} and \citet{T12} revealed that the RV data contain a significant correlated noise component (red noise), which was not treated by any previous astrostatistical methods, as they assumed that the measurement errors were statistically independent, i.e. the noise was uncorrelated (white). Both studies thoroughly analysed the individual and combined RV sets using a red-noise model. \citet{B13} found three robust planetary signals (b, c, e), but the significance of planet d's signal dropped with the red-noise model and its existence will therefore require independent confirmation. \citet{T12}, on the other hand, identified four clear planets (b-d) based on the two data sets using a red-noise model. Interestingly, the HARPS data alone showed a fifth signal with a period of 190 days, but they concluded against its planetary origin, because it was not detected in the combined RV data sets.

Using submillimetre wavebands of the Herschel Space Observatory, \citet{L12} have spatially resolved a cold debris disk around Gliese 581, extending radially from $25\pm12$ to 60 AU. The debris disk was determined to have an inclination between $30^{\circ} < i < 70^{\circ}$. Assuming that the disk mid-plane and the planetary orbits are co-planar, this range of inclination makes the masses of the planets no more than $\sim1.6$ times their measured minimum masses by RV measurements. The known planets around the star within 0.3 AU cannot dynamically perturb the disk sufficiently over the lifespan of the star (2-8 Gyr), which suggests that another undiscovered planet may exist further out, keeping the comet belt replenished.

Since different statistical methods and noise modeling might lead to different orbital solutions in the future, the debate over the number of planets in the Gliese 581 exoplanetary system is far from over yet. Moreover, Gliese 581 is still the target of exoplanet surveys and the addition of new high precision RV data, additional or already discovered periodic signals may become statistically significant and prove to be of planetary origin. Nevertheless, dynamical stability of current models is important to investigate, especially in the light of the debate over the orbital eccentricities of the planets.

Here we will present a numerical investigation into the dynamical stability of the Gliese 581 system, with special focus on stability as a function of eccentricity. Apart from how the eccentricities were treated, the orbital parameters and planetary masses did not change considerably after the 4-planet solution of \citet{V12}, so we look into the long-term dynamical stability of this RV fit. From the initially circular orbits we expect that this model is the most stable dynamically, therefore we search for stable orbits for a fifth planetary body in the inner part of the system, inside the orbit of planet d. (Here, we exclude the region of long-period orbits, such as planet f in the 6-planet model of \citet{V10} would have. Due to the long distance, the inner planets would not influence their stability; conversely, their low masses would not affect considerably the stability of the inner orbits.) We will see that there are stable regions between planet c and d for a fifth planet in the system, and this region corresponds to the liquid water habitable zone of the star. Given this potential, we investigate the long-term stability of the 5-planet model of \citet{V12}, which includes a low-mass planet (the unconfirmed planet g) in this zone. The main objective of our study is to set dynamical constraints on the orbits of these RV fits. Using stability maps established from well-known stability indicators, we will determine the limits of the orbital eccentricities of the planets.

\section{Dynamical models and methods}

We investigated the dynamical stability of the Gliese 581 planetary system by using numerical integrations of the planetary orbits. The model parameters were taken from the 4-planet and 5-planet fits of \citet{V12}, these are presented in Table \ref{tab-4planet} and \ref{tab-5planet} for reference. For the mass of Gliese 581 we took $0.31 M_{\odot}$
. We made the assumption of coplanarity of all orbits, and in case of the circular 4-planet model and with the addition of a fifth test planet, we checked the dynamical stability for different orbital inclinations (to the line of sight), otherwise $\sin i = 1$ (an edge-on system) was assumed. We confined our study area between 0.01 and 0.41 AU. The stability of the following models were investigated:
\begin{enumerate}
\item 5-body problem consisting of the star and the 4 planets (b-e) to check the long-term stability of the 4-planet system,
\item restricted 6-body problem consisting of the star, 4 planets (b-e) and a massless fifth planet to search for stable regions for an additional planetary body,
\item 6-body problem consisting of the star and 5 planets to verify the dynamical stability of the 5-planet system including planet g as presumed by \citet{V12}.
\end{enumerate}
The numerical integrations were performed using the method of Lie-integration with an adaptive step-size, which is a very fast and precise integration method due to the recurrence of the Lie-terms \citep{hanslmeier84, pal07}. In order to characterize the stability of the models, we used the Lyapunov characteristic indicator (LCI), the relative Lyapunov indicator (RLI), and the maximum eccentricity method (MEM). The LCI, the finite time approximation of the maximal Lyapunov Exponent, is a well known chaos indicator of a dynamical system, it estimates the exponential divergence rate of infinitesimally close trajectories in the phase space. The RLI is the difference of the Lyapunov indicators of two very close trajectories and provides an effective tool for detection of chaotic behavior during short integration times \citep{sandor04}. The MEM provides information about the evolution of the orbit during the integration time through the largest value of the eccentricity of the planets and indicates close encounters and escapes from the region of motion as well \citep[e.g.][]{dvorak03, suli05, nagy06}. While the chaos indicators LCI and RLI characterize the structure of the phase space, dynamical stability is explicitly described by the maximum eccentricity. Since chaos does not necessarily mean dynamical instability, rather a sensitivity to initial conditions, we perform long-term integrations of the 4 and 5-planet models as well.

\section{Stability of the 4-planet system}

\begin{table}
\caption{Astrocentric, circular, non-interacting 4-planet orbital model of \citet{V12}. $P$: orbital period, $M_{min} = m \sin{i}$, where $i$ is the orbital inclination, $a$: semi-major axis, $e$: eccentricity, $l$: longitude of the periastron.}
\label{tab-4planet}
\begin{tabular}{lccccc}
\hline
\it Planet & \it P & $M_{min}$  & \it a & \it e & \it l \cr
 & (days) & $(M_{\oplus})$  & (AU) & & (deg) \cr
\hline
\it e &	3.15 & 1.84 & 0.028 & 0 & 138.5 \cr
\it b	&	5.37 & 15.98 & 0.04 & 0 & 338.9 \cr
\it c	&	12.93 & 5.4 & 0.073 & 0 & 175.2 \cr
\it d*	&	66.71 & 5.25 & 0.22 & 0 & 235.8 \cr
\hline
* planet candidate
\end{tabular}
\end{table}

\subsection{Basic stability of the 4-planet model}

First, we investigated the stability of the circular 4-planet model of \citet{V12} in detail, which was found to be stable for at least 100 kyrs by the authors. The analysis of RV variations is able to constrain the mass ($M$) of exoplanets by a lower limit, since only the quantity $M \sin i$ is determined, where $i$ is the unknown orbital inclination. Below a given value of $i$ - or, equivalently above given masses for the planets - we expect the system to become unstable as the dynamical interactions can increase with lower inclinations, i. e. higher planetary masses. Although the debris disk around the star narrows down the inclination to $30^{\circ} < i < 70^{\circ}$ possibly for the planets as well \citep{L12}, this range is not necessarily where the planetary orbits lie. Therefore, to check for which range of inclination the system has this basic stability, we integrated the circular 4-planet model (Table \ref{tab-4planet}) over 1 Myr, varying the inclination by $\triangle i = 5^{\circ}$ from $i=90^{\circ}$ (edge on) to $i=5^{\circ}$ (almost pole on). 

In all cases, the variations of the semi-major axes and eccentricities of the four planets are not significant. The initially circular 4-planet system remains stable for 1 Myr down to $i=5^{\circ}$ (almost pole-on) and co-planar orbits. Dynamical stability with $i \geq 5^{\circ}$ suggests that the mass of each planet can be $\sim12.5$ times of its minimum mass. For Gliese 581 e, b, c, and d those limits are 21, 183, 62 and 60.2 $M_{\oplus}$.

\subsection{Stability of planet e}
\label{stability-e}

The low mass planet e has little effect on the stability of the more massive planets in the system, but an increase in its eccentricity quickly leads to orbital crossings with planet b or ejection from the system \citep[as it was already pointed out by][]{M09, V12}. To characterize planet e's orbit in the circular 4-planet model, we carried out integrations with the following grid of initial semi-major axis and eccentricity: $0.01 \leq a \leq 0.041$ AU with steps $\triangle a = 0.003$ AU and $0 \leq e \leq 1$ with steps $\triangle e = 0.01$, for a time period of 5000 $P_e$, where $P_e$ is the orbital period of planet e.

\begin{figure}
\includegraphics[width=8.4cm]{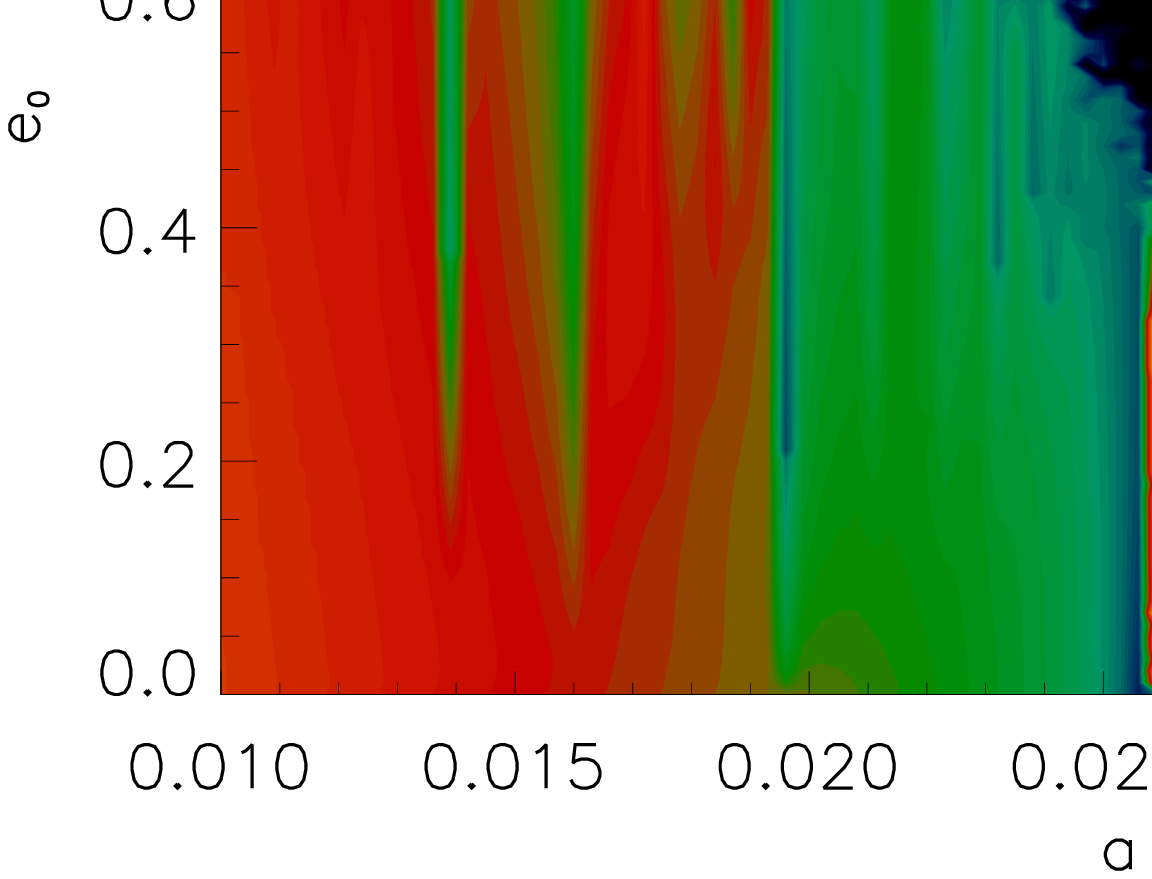}
\caption{a-e stability map of planet e, its maximum eccentricity is plotted as a function of semi-major axis and eccentricity (with a starting value of 0). Yellow regions are stable, black regions denote escaping orbits.}
\label{ME-e}
\end{figure}

The maximum eccentricity that planet e reached over the time of the integration is plotted on the $a-e$ stability map in Figure~\ref{ME-e}. Planet e's orbit is stable (small $e_{max}$), although the vicinity around its initial $a_{e}=0.028, e_{e}=0$ values represents a stable island on the map, so relatively small perturbations could lead to instability and thus collision or ejection. Having varied only the eccentricity $e_{e}$, the LCI became suddenly high above $e_{e}=0.22$ (data not shown). This indicates chaos, and also points to the highest eccentricity for planet e's orbit to remain stable.

We also performed integrations with different starting positions ($l$) for planet e varying $l$ by $2^{\circ}$ between $l=0-360^{\circ}$. The LCI values (data not shown) remain low during the time of the integration for each case, therefore planet e's stability does not depend on its orbital position in the model.

\section{Stability of the 5-planet system}

\subsection{Stable regions for a fifth planet}

\begin{figure*}
\centering
\begin{subfigure}[]
\centering
\includegraphics[width=11cm]{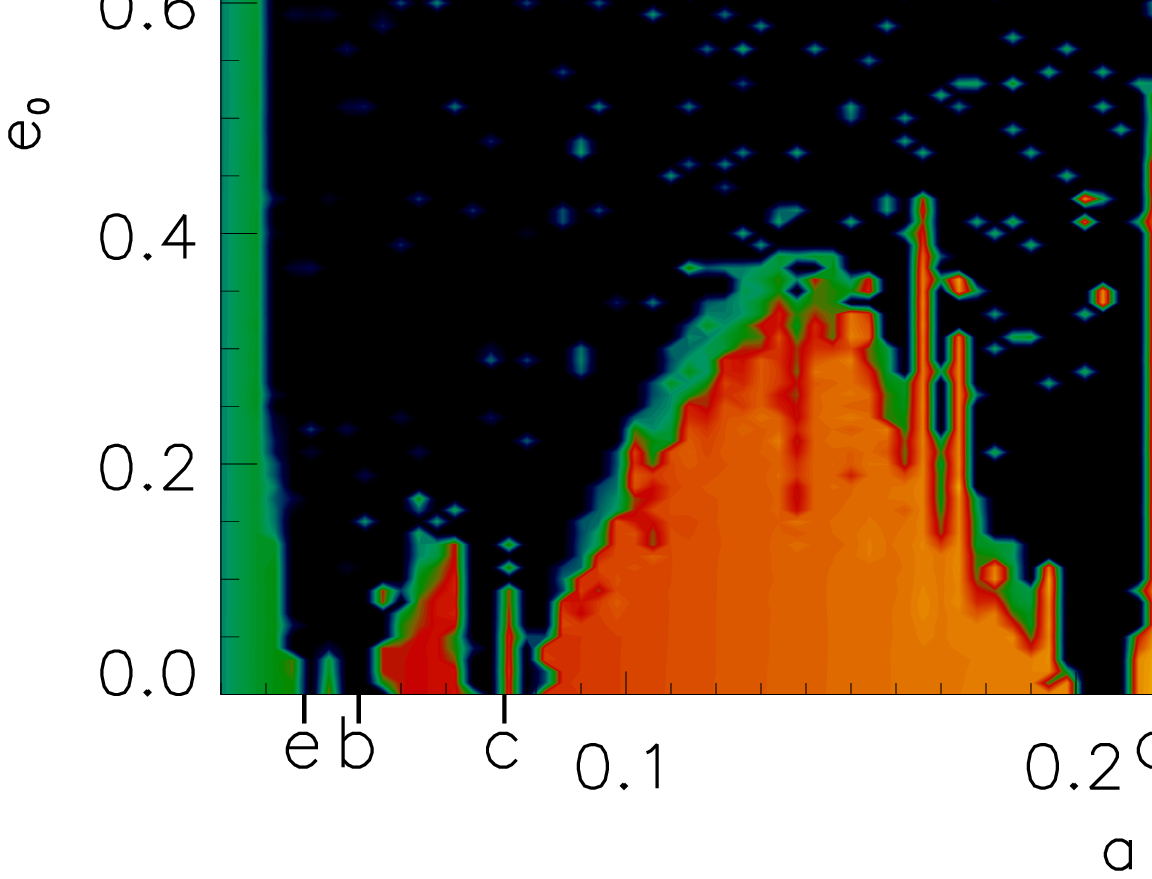}
\end{subfigure}
\begin{subfigure}[]
\centering
\includegraphics[width=11cm]{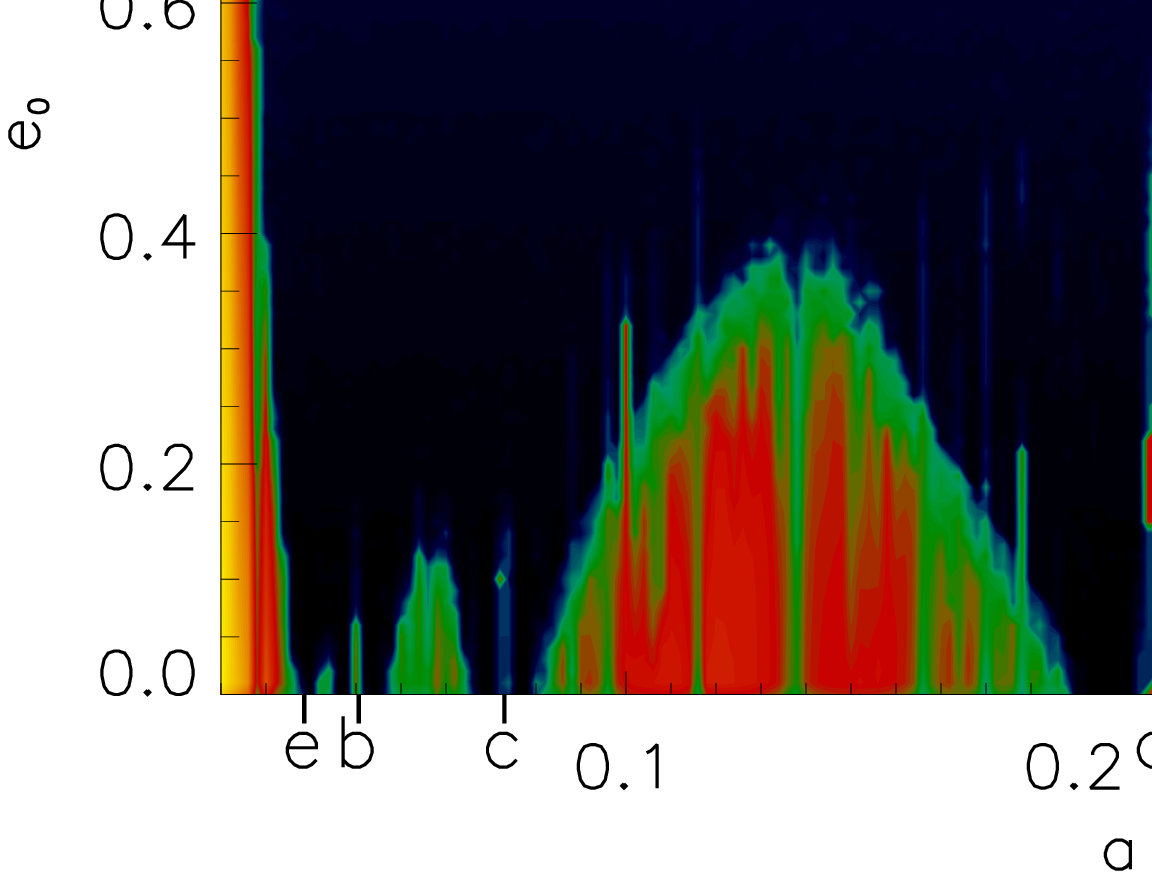}
\end{subfigure}
\caption{(a) LCI values of a fifth massless test planet for different initial $a, e$ pairs. Stable regions are marked as yellow according to the LCI. (b) Maximum eccentricity of a fifth massless test planet. Each $a, e$ point on the map represents an average of 8 maximum eccentricities that the test planet reached over the time of the integrations and started at 8 different starting positions in the model. On top, the main mean-motion resonances are indicated between the fifth test planet and planet c or d. The parameters of these resonances are listed in Table \ref{tab-resonances}. The location of the four planets b-e are indicated at the bottom of each plot.}
\label{LCI-ME-5th-ae}
\end{figure*}

\begin{figure}
\includegraphics[width=8.4cm]{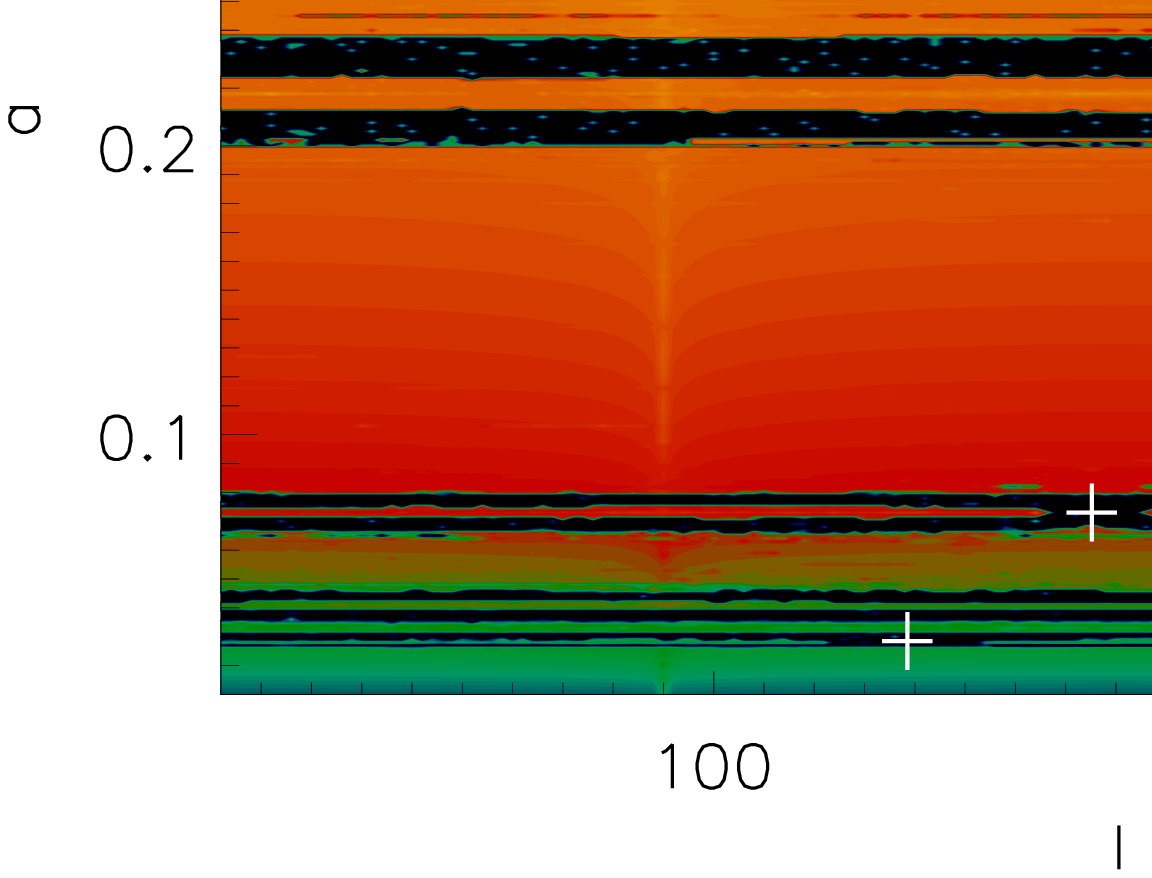}
\caption{LCI values of a fifth massless test planet for different initial semi-major axis ($a$) and starting position ($l$). Chaotic orbits are marked with colors towards black, stable orbits towards yellow.}
\label{LCI-5th-la}
\end{figure}

We integrated the 4-planet system with an additional massless body in order to find regions of possible orbital stability for a fifth planet. The test planet's initial parameters were varied, in one set the semi-major axis between $0.01 \leq a \leq 0.41$ AU with steps of $\triangle a = 0.001$ AU and the eccentricity $0 \leq e \leq 1$ with steps of $\triangle e = 0.01$, and in another set for the same semi-major axis range and $e=0$, the starting position $0 \leq l \leq 360^{\circ}$ with steps of $\triangle l = 2^{\circ}$. The time of the integrations was 10000 $P_{test}$ for calculating the LCI, where $P_{test}$ is the orbital period of the test planet. The LCI values of each integration for the grid of different $a, e$ are plotted in Figure~\ref{LCI-ME-5th-ae}.a, for $a, l$ in Figure~\ref{LCI-5th-la}. The $e_{max}$ that the test planet reached over 5000 $P_{test}$ was calculated for the same grid of $a, e$, and for 8 different starting orbital positions with steps $\triangle l = 45^{\circ}$ between $l=0 - 315^{\circ}$. The average of these 8 $e_{max}$ values \citep{nagy06} are represented on one map in Figure~\ref{LCI-ME-5th-ae}.b.

The orbits of a fifth fictitious planet are chaotic in the vicinity of the innermost planets e and b, but extensive stable regions exist between the larger planets c and d, and outside the orbit of planet d, with relatively low eccentricities (Fig. \ref{LCI-ME-5th-ae}.a). The starting position of the fictitious planet only plays a role in its stability around the innermost planets and its orbit is less chaotic in the outer part of the system (Fig. \ref{LCI-5th-la}). The $a-e$ stability map with the maximum eccentricity essentially shows the same boundaries for stable regions, in fact the average $e_{max}$ values confirm that irrespective of the starting position of the test planet, its orbit can be stable in the region between the super-Earth planets and beyond them (Fig. \ref{LCI-ME-5th-ae}.b). However, unstable configurations can be observed in both of these regions on the map (see further details about the habitable zone in section \ref{stabilityHZ}). This is in agreement with the results of \citet{TdS12}, who gave dynamical constraints for the existence of an additional planet between planet c and d.

The integration of the 4-planet model including a fifth test planet was also carried out for different inclinations to the line of sight and for  different degrees of the argument of pericenter ($\omega$). The integrations ran for 5000 $P_{test}$ for inclinations $i = 30, 50, 70^{\circ}$--in the range where the inclination of the debris disk around the star lies; while the argument of pericenter was varied between $0 \leq w \leq 360^{\circ}$ with steps of $\triangle \omega = 90^{\circ}$. Neither of these parameters changed significantly the possible stable regions in the system for a fifth planet based on the maximum eccentricity of the test planet, in fact the resulting $a-e$ stability maps (see Figure S1 in the Supplementary Material online) resemble very closely that in Figure~\ref{LCI-ME-5th-ae}.b in each case.

On the stability maps in Figure~\ref{LCI-ME-5th-ae} and Figure~\ref{LCI-5th-la}, there are small stable islands visible sporadically, where both the $e_{max}$ and the LCI have low values at the end of the integration. These point to mean-motion resonances (MMRs) between the test planet and the other planets. While no MMR could be identified in the inner part of the system (due to the even parametrization of the stability maps), several MMRs were found involving the test planet and the two super-Earth planets around the habitable zone. The main resonances are indicated in Figure~\ref{LCI-ME-5th-ae}.b and explained in Table~\ref{tab-resonances}. MMRs enhance stability and create stability islands in these region, because planet c and d are assumed to have zero or very low eccentricity orbits \citep{yoshikawa89, beust96}. Nevertheless, none of these low order MMRs overlap, which could lead to instability in the planetary system.

\begin{table}
\caption{Main mean-motion orbital resonances between a fifth test planet and planets c or d in the Gliese 581 planetary system indicated on the $a-e$ stability map in Figure~\ref{LCI-ME-5th-ae}. $(p+q)/p$ is equal to mean motion ratio of the two planets in resonance. $q>0$ for inner resonances, $q<0$ for outer resonances.}
\label{tab-resonances}
\begin{tabular}{lcccc}
\hline
\it Resonance & $(p+q)/p$ & \it a & \it p & \it q \cr
 & & (AU) & & \cr
\hline
c1 & 7/8 & 0.0798 & 7 & -1 \cr
c2 & 2/3 & 0.0956 & 2 & -1 \cr
c3 & 5/8 & 0.0998 & 5 & -3 \cr
c4 & 1/2 & 0.116  & 1 & -1 \cr
c5 & 1/3 & 0.1518 & 1 & -2 \cr
d1 & 7/3 & 0.124 & 3 & 4 \cr
d2 & 2/1 & 0.137 & 1 & 1 \cr
d3 & 5/3 & 0.155 & 3 & 2 \cr
d4 & 3/2 & 0.166 & 2 & 1 \cr
d5 & 7/5 & 0.174 & 5 & 2 \cr
d6 & 4/3 &	0.18 & 3 & 1 \cr
d7 & 5/4 & 0.188 & 4 & 1 \cr
d8 & 6/5 & 0.192 & 5 & 1 \cr
d9 & 7/8 & 0.238 & 7 & -1 \cr
d10 & 3/4 & 0.264 & 3 & -1 \cr
d11 & 2/3 & 0.2855 & 2 & -1 \cr
d11 & 1/2 & 0.346 & 1 & -1 \cr
d13 & 2/5 & 0.4014 & 2 & -3 \cr
\hline
\end{tabular}
\end{table}

\subsection{Stability of the 5-planet model including planet g}

\begin{table}
\caption{Astrocentric, circular, non-interacting 5-planet orbital model of \citet{V12}. $P$: orbital period, $M_{min} = m \sin{i}$, where $i$ is the orbital inclination, $a$: semi-major axis, $e$: eccentricity, $l$: longitude of the periastron.}
\label{tab-5planet}
\begin{tabular}{lccccc}
\hline
\it Planet & \it P & $M_{min}$  & \it a & \it e & \it l \cr
 & (days) & $(M_{\oplus})$  & (AU) & & (deg) \cr
\hline
\it e &	3.15 & 1.86 & 0.028 & 0 & 141.9 \cr
\it b	&	5.37 & 16.0 & 0.04 & 0 & 338.4 \cr
\it c	&	12.93 & 5.3 & 0.073 & 0 & 181.0 \cr
\it g*	&	32.13 & 2.24 & 0.13 & 0 & 55.3 \cr
\it d*	&	66.67 & 5.94 & 0.22 & 0 & 227.3 \cr
\hline
* planet candidate
\end{tabular}
\end{table}

Even though the existence of planet g, the fifth planet candidate in the star's classical liquid water 
habitable zone is debated and might as well be a false periodic signal caused by noise, we investigated its possibility from a dynamical point of view. For this first, the 5-planet model of \citet{V12} (as a reference see Table \ref{tab-5planet}) containing planet g was integrated for 1 Myr to verify its long-term stability. All 5 planets remained stable during the long integration time and their orbital parameters did not change significantly, so this 5-planet solution of the RV data with the 32-day planet g can be considered dynamically stable.

\section{Stability as a function of eccentricity}

\subsection{Planet e, b, c}
\label{e-ebc}

The orbits of the planets in the Gliese 581 planetary system might be elliptic as a consequence of dynamical interactions. Highly eccentric orbits can cause instability though, we derived therefore limits for the eccentricities using $a-e$ stability maps. First, we investigated the situation of the inner planets e, b and c. We have assumed that given its higher mass, planet b is likely to be more stable in comparison to the adjacent planets, so we looked at the stability of planet e for a range of initial $e_e$ and $e_b$, and the stability of planet c for a range of initial $e_b$ and $e_c$ (in this case planet e was left out of the system). The maximum eccentricity was calculated over a time period of 5000 $P_e$ and 5000 $P_c$, respectively. The eccentricities ranged until the elliptical orbit of one planet crossed the other one's circular orbit.

\begin{figure}
\includegraphics[width=8.4cm]{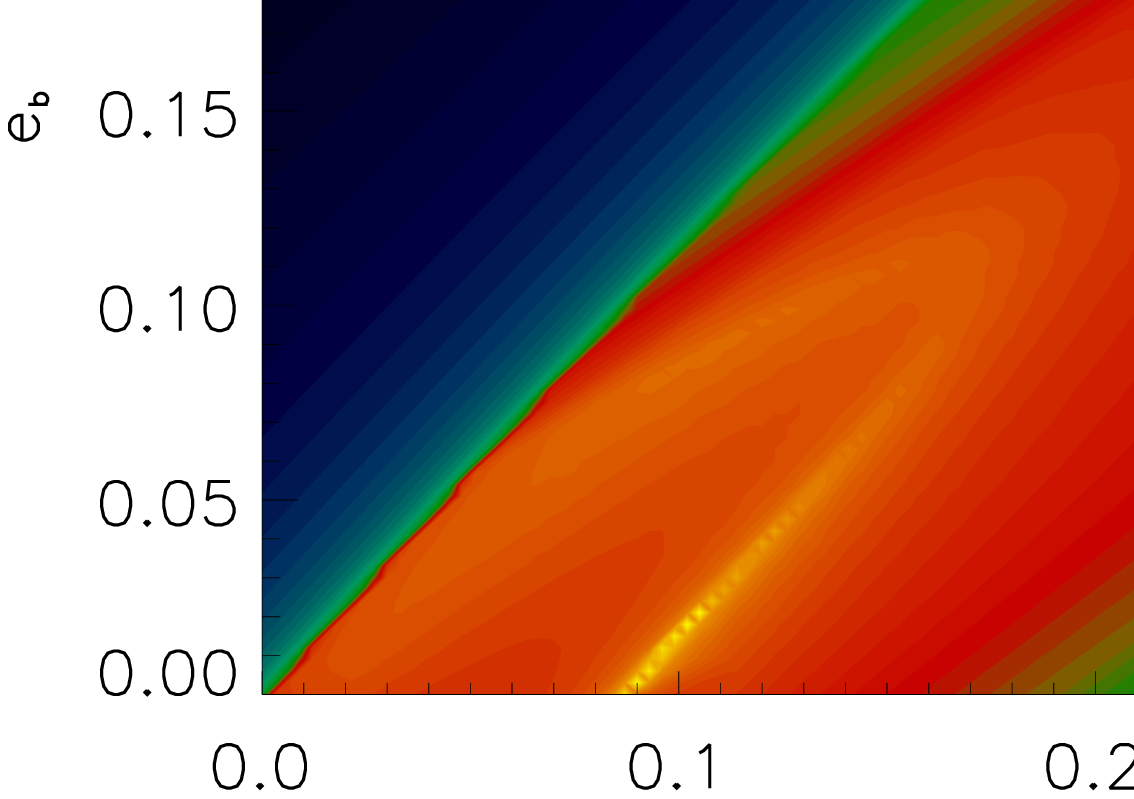}
\caption{Stability of planet e is characterized with its the maximum eccentricity as a function of $e_e$ and $e_b$. Black denotes unstable, whereas yellow stable configurations.}
\label{ME-eb-ee}
\end{figure}

The maximum eccentricity, plotted in Figure~\ref{ME-eb-ee} as a function of initial $e_e$ and $e_b$, indicates that planet e's eccentricity is coupled to the more massive planet b's eccentricity. If $e_b \simeq 0$, the limit of planet e's eccentricity is $e_e \leq 0.18$. If planet b's eccentricity is above zero but relatively low, then planet e remains stable below $e_e \leq 0.2$, which is approximately the same upper limit as the one based on the LCI (in section \ref{stability-e}). Small stable regions exist with eccentricities above $e_e = 0.3$ too (Fig. \ref{ME-eb-ee}), though these isolated configurations are less likely to remain stable for a longer time.

\begin{figure}
\includegraphics[width=8.4cm]{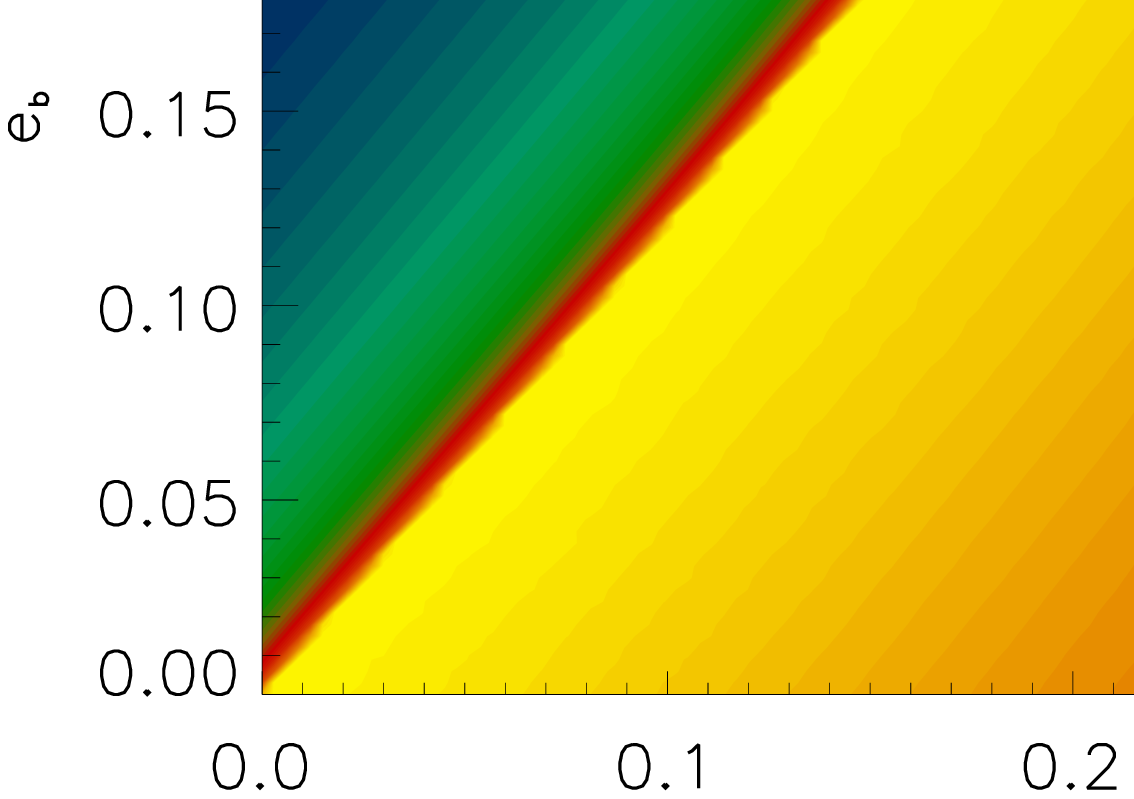}
\caption{Stability of planet c is characterized with its the maximum eccentricity as a function of $e_b$ and $e_c$. Black denotes unstable, whereas yellow stable configurations.}
\label{ME-bc-ee}
\end{figure}

Looking at the stability of planet c as a function of $e_b$ and $e_c$ in Figure~\ref{ME-bc-ee}, it is apparent that planet c puts less constraint on the eccentricity of planet b (which is then mainly given by the smaller planet e). While in case of circular or near-circular orbits for planet b, the upper limit of planet c's eccentricity is $e_c \leq 0.32$.

\subsection{Planet c, d and a planet in the habitable zone}
\label{stabilityHZ}

The stability of the outer planet d is less influenced by planet c because of the large distance between them, but a planet between planet c and d in the habitable zone would put constraints on both planet c and d's eccentricity. For this reason, the stability of planet c and d is investigated with the 32-day planet g in the system or in general with another planet in the habitable zone.

\begin{figure}
\includegraphics[width=8.4cm]{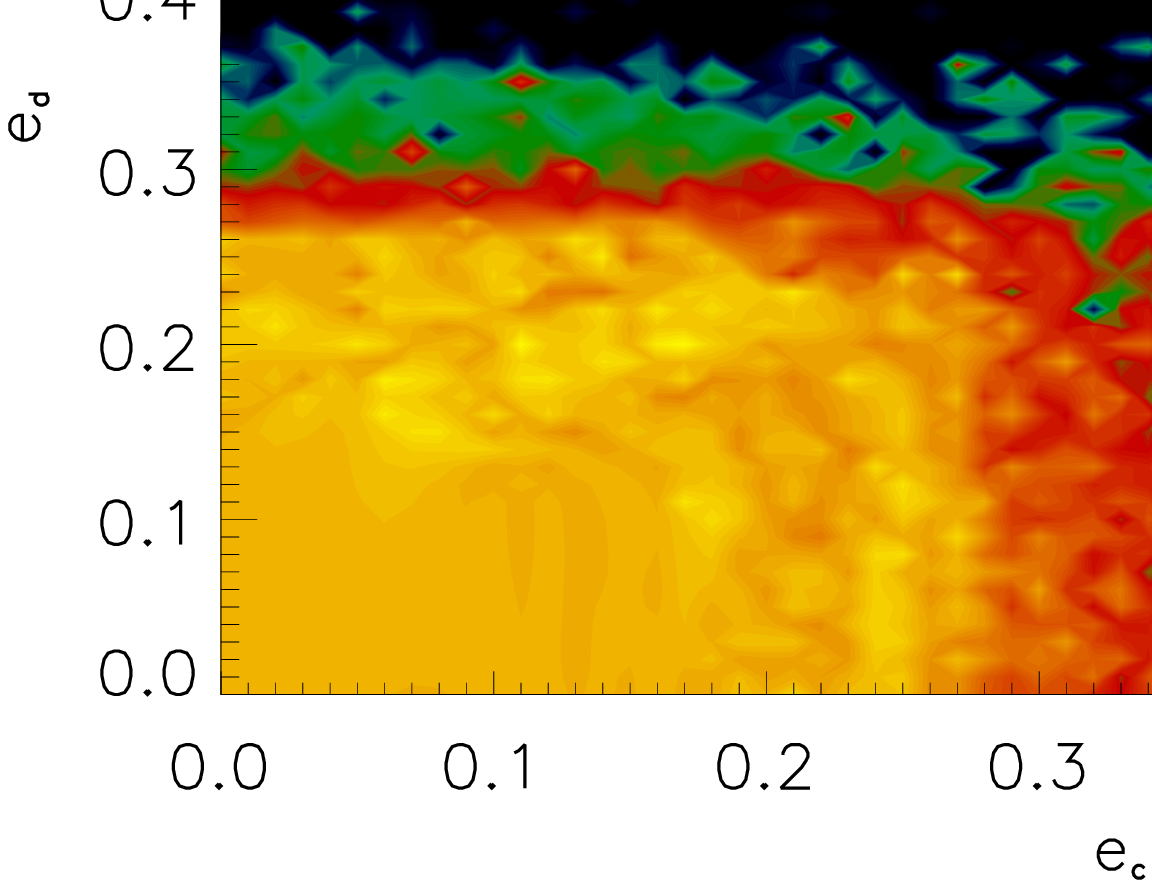}
\caption{LCI values of the orbit of planet g as a function of eccentricity of the adjacent planets c and d. The yellow to red rectangular region indicates stable orbits, initial points in the dark regions result in chaotic motion.}
\label{LCI-g-cd-ee}
\end{figure}

We ran two set of integrations to characterize the stability of planet g's orbit as a function of the adjacent planets' eccentricities. In one set, for a time of 5000 $P_g$, the eccentricity of the adjacent planets c and d were varied until the point when their orbits cross each other. The LCI of planet g in Figure~\ref{LCI-g-cd-ee} shows that planet g's orbit can be considered stable in the rectangular region on the map where the eccentricity of planet c stays below $e_{c} \leq 0.32$ and of planet d below $e_{d} \leq 0.28$. These eccentricity limits mean $r_{per}=0.049$, $r_{ap}=0.096$ AU pericenter and apocenter distances for planet c, and $r_{per}=0.158$, $r_{ap}=0.282$ AU for planet d. The apocenter of the orbit of planet c leaves a little bit more space for planet g (at 0.13 AU) than the pericenter of the orbit of planet d. Interestingly, however, planet c then causes instability in the orbit of planet g even in case of moderately close encounters, even though planet c's mass is lower than that of planet d. The reason behind this may be that planet c's orbital period is about the fifth of the period of planet d, thus planet c simply approaches planet g more frequently, causing more perturbation. Referring back to the previous section, planet c's eccentricity as high as $e_c \simeq 0.32$ would not actually be the only cause of the instability in planet g's orbit. When $e_c \simeq 0.32$, the adjacent, more massive planet b already considerably affects the stability of planet c (Fig. \ref{ME-bc-ee}) and this is what may ultimately lead to the instability of planet g.

\begin{figure}
\includegraphics[width=8.4cm]{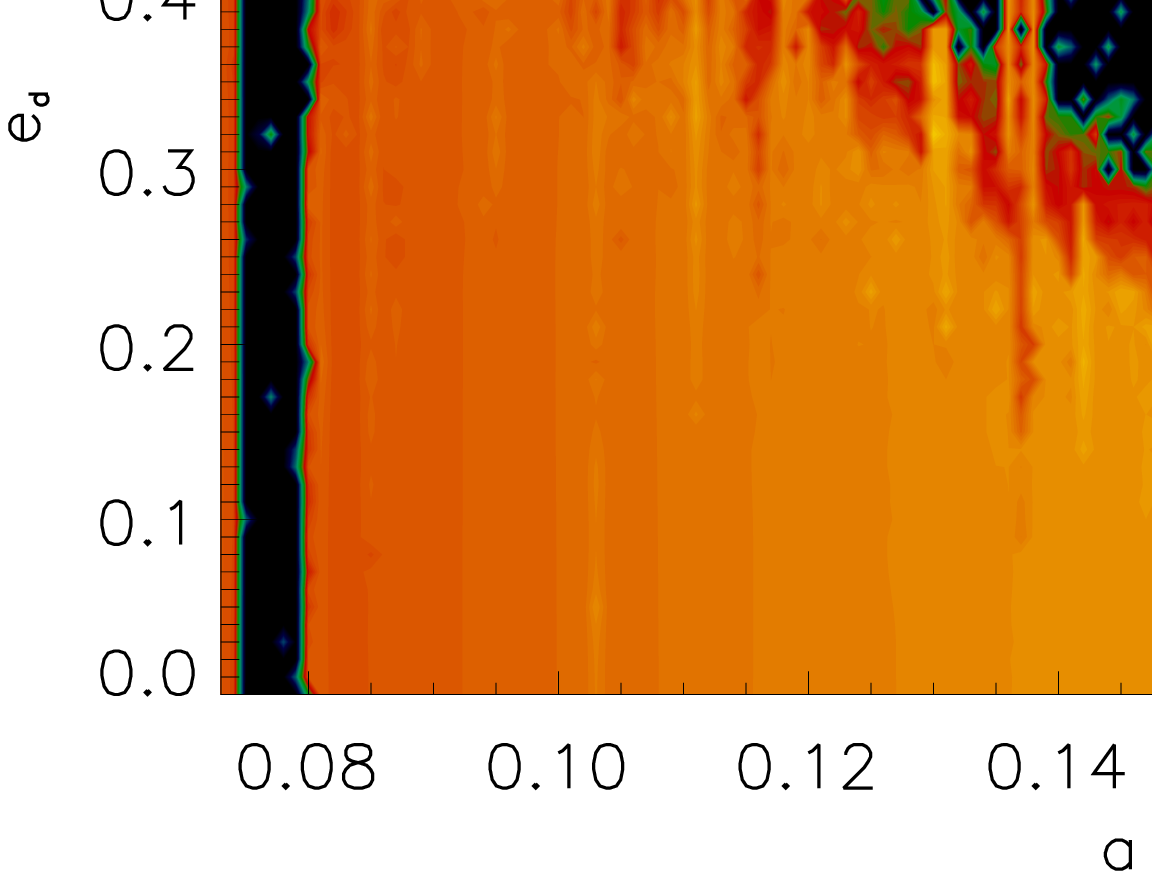}
\caption{Stability of planet g as a function of planet d's eccentricity and the semi-major axis. Yellow marks the stable, black end of the color bar the chaotic orbits according to the LCI.}
\label{LCI-g-d-ae}
\end{figure}

Since planet d's real eccentricity could be greater than zero based on the RV fits, as well as the true orbital period and thus semi-major axis of planet g can be different, in another set of integrations, also for a time of 5000 $P_g$, we varied $a_g$ in the range between the neighbouring planets, and $e_d$ until planet d's orbit would cross the orbit of planet c. Not surprisingly, as Figure~\ref{LCI-g-d-ae} shows, the more elliptic the orbit of planet d, the further away planet g has to be from it in order to remain stable. At the largest distance, planet g's orbit would be stable in case of $e_d \approx 0.55$, but such an elliptic orbit would leave only a very narrow region in the system for the fifth planet. Eccentricity as high as $e_{d}\simeq0.4$ that was suggested in previous models for planet d \citep{M09, G11, T11, F11}, would limit planet g's orbit to be in a narrow region between $\sim 0.08-0.12$ AU (Fig. \ref{LCI-g-d-ae}). As \citet{TdS12} has already pointed it out, stability of planet g is therefore strongly dependent on the eccentricity of the adjacent planet d. More generally, based on the map in Figure~\ref{LCI-g-d-ae}, it can be stated that this dependence is true for the stability of a low-mass planet orbiting in the habitable zone between planet c and d. Also, several two or three-body mean motion resonances between planet g and planet c or d were determined in the region (Fig. \ref{LCI-ME-5th-ae}.b), which are indicated here by the small vertical chaotic regions observed on the map in Figure~\ref{LCI-g-d-ae}.

\begin{figure}
\includegraphics[width=8.4cm]{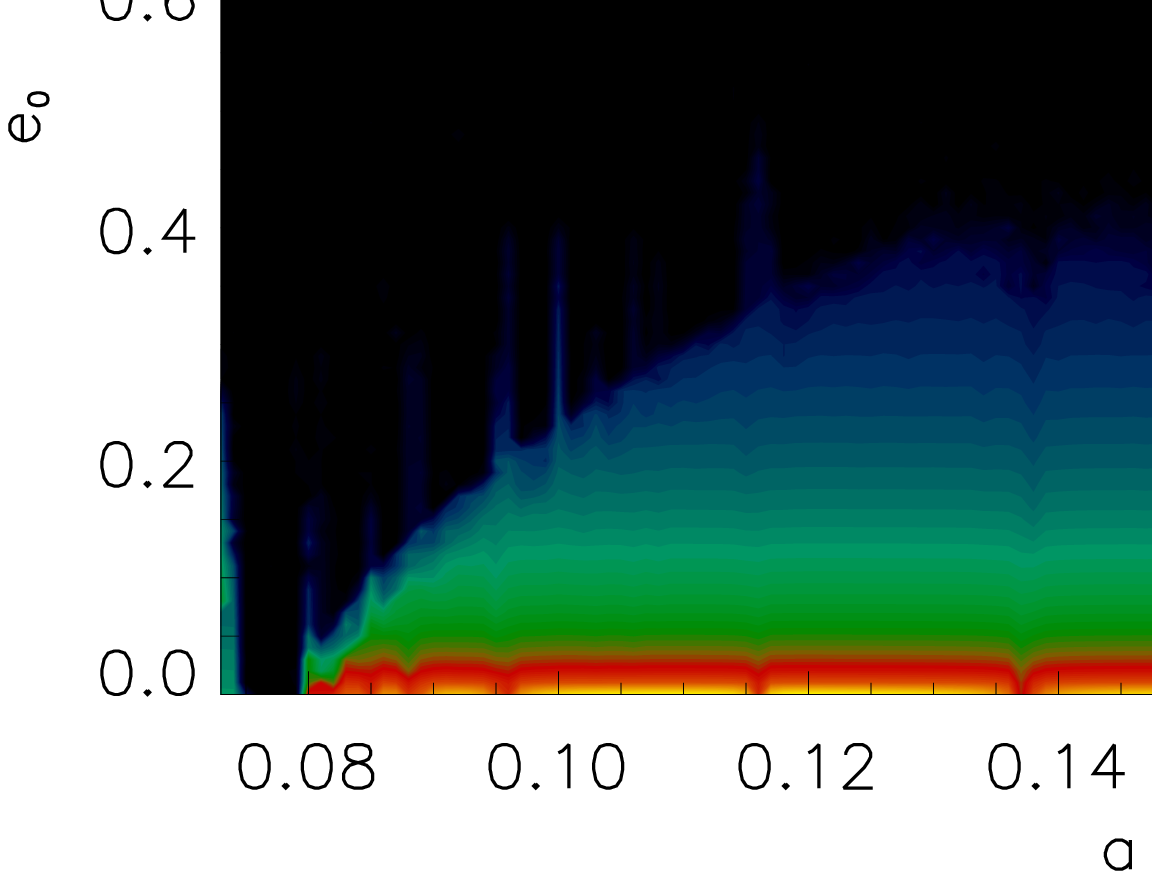}
\caption{$a-e$ stability map of a massless test planet in the region between Gliese 581 c and d. On the map, the average RLI is plotted, which comes from 8 integrations starting with 8 different starting orbital positions of the test planet. Yellow marks the stable, black end of the color bar the chaotic orbits according to the RLI values.}
\label{RLI-5th-ae-l}
\end{figure}

Finally, we investigated the general case of a planet in the habitable zone by checking for which range of eccentricity a planet would remain stable here. We integrated the all-circular 4-planet model with the addition of a massless test planet and varied its eccentricity. The integration ran for a time of 5000 $P_{test}$ and for 8 different starting orbital positions with steps $\triangle l = 45^{\circ}$ between $l=0 - 315^{\circ}$. This time we calculated the RLI of the test planet as a function of the semi-major axis and eccentricity. The average of these 8 RLI values is plotted on the $a-e$ stability map in Figure~\ref{RLI-5th-ae-l}. In general, the dynamical stability of another planet between planet c and d greatly depends on the eccentricity, and a planet would be only stable in this region on circular or nearly circular orbits. Even this very small stable region close to zero eccentricity is not continuous, probably as a consequence of the above mentioned MMRs between the fifth planet and planet c and d.

\section{Conclusions}

In the process of fitting multi-planet models to precise RV data, it is a question of choice whether to allow the orbital eccentricities to float freely or to hold them fixed at zero. Comparing the 4-planet models of \citet{M09, F11} and \citet{V12} in regards to their long-term dynamical stability, eccentricity plays a great role \citep[see also the discussion of][]{V12}. When eccentricities are fixed at zero, the degree of freedom in fitting the RV curve is reduced, and such a model is more likely to describe a stable system. As the case of the Gliese 581 planetary system shows, freely floating eccentricities can lead to solutions that are unstable and therefore unphysical despite providing good fits to the RV data. The large uncertainty of eccentricities should warn us that there is not enough RV data yet to be able to constrain orbital parameters in a Keplerian model. Forced circular orbits in the RV fits might be too strong a restriction though.

Long-term numerical integrations show that the all-circular 4-planet model of \citet{V12} remains dynamically stable for 1 Myr, even for an inclination $i = 5^{\circ}$, i. e. with $\sim12.5$ times higher planetary masses than the minimum mass. In this model, the innermost planet e's orbit remains stable during the integration time, however the stable region on the $a-e$ plane around it is rather small.

A fifth planetary body in the 4-planet system could have a stable orbit between the two super-Earth planets c and d, and there is a stable region for an additional planet beyond the orbit of planet d. The 5-planet model of \citet{V12}, which includes the disputed planet g, is also dynamically stable on a longer timescale. The existence of the low-mass planet g in the habitable zone is supported from the dynamical point of view and it can be considered stable for a wider range of orbital configurations. 

While the models of \citet{V12} with initial circular orbits are dynamically stable over a longer time period, the true orbits might be elliptic as a consequence of dynamical interactions between the planets. Our simulations put dynamical constraints on the eccentricity of the planets, and these limits can in turn be used to help find the best stable model to fit the RV measurements. The Neptune-sized planet, Gliese 581 b given its high mass greatly influences the stability of the adjacent planets e and c, and the eccentricity of these planets is therefore dependent on planet b's eccentricity. In case of $e_b \simeq 0$, dynamical stability of the inner planets is only ensured, if planet e's eccentricity stays below $e_e \leq 0.18$, and planet c's eccentricity below $e_c \leq 0.32$. In case planet b's orbit is slightly elliptic, the orbit of planet e and c has to be elliptic as well for the inner planets to remain stable. The outermost Gliese 581 d is far away from the neighboring planet c to considerably influence its stability, however, the existence of a planet between the two super-Earth planets puts constraint on the eccentricity of planet d. If a low-mass planet like the 32-day planet g orbits in the habitable zone, then planet d's eccentricity cannot be larger than $\sim0.28$. On the other hand, the stability of a planet like planet g is strongly dependent on the eccentricity of the outermost planet d. This is also true in general for an additional planet orbiting in the habitable zone between planet c and d, which would likely only be stable on circular or near-circular orbit. (Note that eccentricity limits are valid for an inclination of $i = 90^{\circ}$. If $i > 90^{\circ}$, the limits of the eccentricities decrease as well.)

\section*{Acknowledgements}
We greatly appreciate the help of William Brocas with the manuscript and we thank the anonymous reviewers for providing constructive comments.

\label{lastpage}

\end{document}